\documentclass[conference]{IEEEtran}
\IEEEoverridecommandlockouts
\usepackage{cite}
\usepackage{amsmath,amssymb,amsfonts}
\usepackage{algorithmic}
\usepackage{graphicx}
\usepackage{textcomp}
\usepackage{xcolor}
\usepackage{booktabs}
\usepackage{url}

\def\BibTeX{{\rm B\kern-.05em{\sc i\kern-.025em b}\kern-.08em
    T\kern-.1667em\lower.7ex\hbox{E}\kern-.125emX}}

\bstctlcite{IEEEexample:BSTcontrol}

\begin{document}

\title{Characterizing Software Aging in GPU-Based LLM Serving Systems
\thanks{This work was conducted at the College of Computing and Informatics, University of North Carolina at Charlotte.}
}

\author{\IEEEauthorblockN{Domenico Cotroneo \quad Bojan Cukic}
\IEEEauthorblockA{\textit{College of Computing and Informatics} \\
\textit{University of North Carolina at Charlotte}\\
Charlotte, NC, USA \\
\{d.cotroneo, bojan.cukic\}@charlotte.edu}
}

\maketitle

\begin{abstract}
This paper proposes an empirical methodology to study software aging in GPU-based LLM serving systems. Traditional aging studies focus on CPU-centric software with relatively regular workloads; LLM serving is different, spanning a Python host and a CUDA device, handling requests whose cost varies by orders of magnitude, and relying on rapidly evolving software stacks. We run a 216-hour campaign across six co-located deployments under identical stress conditions, monitor host, device, and client metrics in parallel, and apply a statistical pipeline that accounts for autocorrelation and multiple testing. Our results reveal statistically significant memory aging in all deployments, with leak rates strongly dependent on the serving runtime and deployment configuration. Beyond these findings, we provide a reproducible framework that opens a research direction at the intersection of the software aging and rejuvenation and LLM serving communities.
\end{abstract}

\begin{IEEEkeywords}
software aging, large language models,
LLM serving, vLLM, NVIDIA Triton, empirical study, GPU
reliability
\end{IEEEkeywords}

\section{Introduction}
\label{sec:intro}
Large Language Models (LLMs) have moved from research artifacts
to production infrastructure in a few years. Engines like vLLM
and NVIDIA Triton run continuously on expensive GPU hardware,
and are expected to deliver predictable performance over days
or months. Yet the literature on these engines focuses almost
entirely on peak metrics measured in
minutes~\cite{kwon2023vllm,zhang2025jenga,agarwal2024symphony,cheng2024kunserve}.
What happens to them over longer windows is mostly unknown.

For nearly three decades, the software aging and rejuvenation
(SAR) community has studied how long-running systems degrade
over time through memory leaks, resource exhaustion, and
performance drift~\cite{trivedi2004rejuvenation,cotroneo2011survey}.
These effects have been characterized in databases,
virtualization, operating systems, and blockchain
platforms~\cite{wada2022unikernel,cotroneo2022android,soeda2024hyperledger}.
The SAR community has not yet engaged with LLM serving on the
GPU.

To the best of our knowledge, this setting is different from anything studied before. Traditional aging
studies target CPU-bound systems with mature allocators and workloads of
regular cost. LLM serving on the GPU breaks all three assumptions. The
runtime spans two coupled environments: a Python orchestration layer on the
host, and a CUDA inference kernel on the device, with state held on both
sides. A single request can consume from hundreds to thousands of tokens,
with multi-order-of-magnitude variation in cost. And the engines themselves
are young codebases built around optimizations (paged KV-cache, continuous
batching, async schedulers) that have no direct precedent in the systems
where aging was historically characterized.

We address this gap with an empirical campaign on three NVIDIA L40S GPUs in
one host, each serving Qwen2.5-7B-Instruct. We run the three engines, vLLM
standalone, Triton-wrapped vLLM, and a naive PyTorch + HuggingFace baseline,
side by side under the same Poisson stress workload for 36 hours, so that
they contend for the same host CPU, memory, and I/O. This reproduces a
realistic multi-tenant deployment, where several engines are co-located on
one machine rather than measured in isolation. But the three engines differ
in more than one way at the same time, so comparing them alone cannot tell
us which difference causes the aging. We therefore add two targeted
ablations, each isolating one factor. The first re-runs the naive baseline
at low load, separating the effect of the saturated workload from the effect
of the framework itself. The second is a 2x2 factorial that crosses the two
vLLM engine generations (the legacy V0 engine and the redesigned V1 engine,
introduced in the project's recent architectural transition) with the two
deployment modes (standalone server and Triton-wrapped server), separating
engine generation from hosting layer. We monitor 34 indicators per run,
covering system, process, GPU, and client-side metrics.

Our statistical pipeline is designed for autocorrelated data:
non-parametric trend detection with a serial-correlation
correction~\cite{mann1945,hamed1998modified}, non-parametric slope
estimation with exact confidence
intervals~\cite{sen1968,theil1950,hollander1999nonparametric}, and
false-discovery-rate control~\cite{benjamini1995fdr}. A trend is
significant only when the test and the confidence interval agree.

The contribution is three-fold. First, we provide the
evidence that software aging exists in modern LLM serving on the
GPU: over 36 hours, every deployment develops a slow but
statistically significant growth of process-private memory, tens
of kilobytes per hour. Second, and against intuition, the leak
does not track how optimized an engine is. The naive baseline is
not the cleanest; vLLM standalone is, while the heaviest by far
is Triton wrapping the legacy V0 engine. Since all of them share
the same inference path, the leak cannot live there: it lives in
the surrounding runtime, and is a property of the full
deployment, not of any single component. Third, the deployments
do not age the same way: only the Triton + V0 combination grows
in discrete steps, resident and virtual memory rising together,
while the others drift continuously. Throughout we rely on a
reproducible methodology: rate-sweep calibration, open-loop
stress, autocorrelation-aware detection, multiple-testing
correction, and controlled ablation.

We acknowledge the preliminary nature of this study, and we
embrace it: the scope is narrow by design, the method is not. The narrow part is mainly time: each configuration runs for 36 hours, and effects that build up more slowly may not show yet. Within that window the analysis is systematic and
supports a clear conclusion, that software aging is present in
modern LLM serving on the GPU. A few further limits bound how far
the findings reach, without touching their validity. The workload is controlled and repeatable, but it does not reproduce the bursty traffic of real deployments.  We return to these in Section~\ref{sec:threats}.

\section{Background and Related Work}
\label{sec:related}

\subsection{How modern LLM serving engines work}
\label{sec:bg-serving}

A serving engine is a long-running process that loads a trained
language model into GPU memory once at startup and then answers
incoming requests by generating text. Each request consists of
an input prompt (a sequence of tokens) and a budget of tokens
to produce as output. Generation is autoregressive: the model
emits one token at a time, and each new token depends on all
previous ones through an internal data structure called the
key-value cache (KV-cache). The KV-cache for an active request
can occupy hundreds of megabytes of GPU memory and grows for
the entire duration of that request.

The main engineering challenge is therefore memory management
on the GPU. Modern engines address it with two key
optimizations. \emph{Continuous batching} interleaves the
generation of multiple in-flight requests within a single
forward pass, so that the GPU is never idle waiting for one
slow request to finish. \emph{Paged attention}~\cite{kwon2023vllm}
manages the KV-cache in small fixed-size blocks, similar to
virtual memory paging in operating systems, eliminating
fragmentation under heterogeneous request sizes. vLLM is the
reference implementation of paged attention and the de-facto
state of the art in open-source serving.

Within vLLM, two generations of the internal engine coexist.
The original \emph{V0} engine~\cite{kwon2023vllm} is a
synchronous implementation, mature and used as the default in
early production deployments. The newer \emph{V1}
engine~\cite{pytorch-vllm-v1} restructures the
scheduler and the request lifecycle around asynchronous tasks;
it is the default in current upstream releases but remains
marked as experimental, and can be reverted to V0 via an
environment variable. The two engines share the same paged-
attention algorithm and the same external API; they differ in
how requests are scheduled and batched and in how per-request
resources are acquired and released. Whether this choice has
reliability consequences over long-running operation is, to our
knowledge, an open question.

The three configurations we study span the deployment spectrum.
\emph{vLLM standalone} runs paged attention with its own minimal
HTTP server, on the V1 engine in current upstream releases.
\emph{NVIDIA Triton} is a more general inference server that
hosts vLLM as one of its backends, adding a request queue, a
dynamic batcher, an OpenAI-compatible frontend, and other
production features around the same paged-attention engine.
Triton spawns vLLM as a managed subprocess and incorporates a
specific NVIDIA-patched release that runs the V0 engine; the
two vLLM configurations therefore differ not only in
orchestration but also in engine generation. By contrast, a
\emph{naive PyTorch + HuggingFace} server loads the model
directly through the HuggingFace \texttt{transformers} library
and serves one request at a time, without continuous batching,
without paged attention, and without a custom memory manager.
These three configurations span the spectrum from "raw research
code" to "production-grade orchestration", which is precisely
the range across which we characterize aging behavior.

\subsection{Software aging and rejuvenation}
\label{sec:related-sar}

Software aging and rejuvenation (SAR) has been studied for nearly
three decades, since Trivedi and
Vaidyanathan~\cite{trivedi2004rejuvenation} formalized aging as
time-dependent degradation amenable to stochastic modeling. The
field has matured around a methodological core that combines the
Mann-Kendall trend test~\cite{mann1945,kendall1948} with Sen's
slope estimator~\cite{sen1968,theil1950,hollander1999nonparametric}.
Empirical case studies span operating
systems, virtualization~\cite{cotroneo2013JVM}, databases,
blockchain platforms~\cite{soeda2024hyperledger},
unikernels~\cite{wada2022unikernel}, and mobile operating
systems~\cite{cotroneo2022android,cotroneo2011survey}. Notably,
autocorrelation and multiple-testing corrections, both
well-established practices, are rarely
applied~\cite{hamed1998modified,benjamini1995fdr}. Aging-related
bug prediction has more recently been approached with deep
learning over program graphs~\cite{zhang2024sgt}, extending the
methodological repertoire of the field beyond pure time-series
analysis.

The aging behavior of AI and LLM systems has received attention
only very recently. Early work investigated object detection and
image classification systems on edge servers, where
Watanabe~\textit{et al.}~\cite{watanabe2023edgeaging} reported
software aging signatures in real-time object detection
pipelines under sustained operation. Three more recent works
target LLM systems specifically.
Santos~\textit{et al.}~\cite{santos2026memory} report memory
aging during LLM-based inference across three open-source models
(Pythia, OPT, GPT-Neo), but restrict the experimental setup to
CPU resources. Santos, Andrade, and
Natella~\cite{santos2025investigating} investigate aging in
applications \emph{generated} by LLMs rather than in serving
engines themselves, showing through 50-hour experiments that
aging propagates into LLM-authored code. Moura
Silva~\textit{et al.}~\cite{moura2025adaptive} propose an
adaptive ML-based detector for software aging under
workload-shift conditions, again on CPU systems. None of these
works addresses the GPU-resident, framework-mediated workload
profile of modern LLM serving engines.

In parallel, the systems community has extensively studied LLM
serving from a performance angle:
PagedAttention~\cite{kwon2023vllm} and a family of memory
management
strategies~\cite{prabhu2024vattention,zhang2025jenga,agarwal2024symphony,cheng2024kunserve}
treat fragmentation and KV-cache pressure as architectural
problems to be eliminated by design, not as time-accumulating
signatures to be detected through long-window analysis.

To our knowledge, this paper is the first to apply software
aging methodology to GPU-based LLM serving engines with
multi-engine comparison, autocorrelation-aware trend detection,
and false-discovery-rate-corrected multiple-testing control.

\section{Methodology}
\label{sec:methodology}
\begin{figure}[t]
\centering
\includegraphics[width=\linewidth]{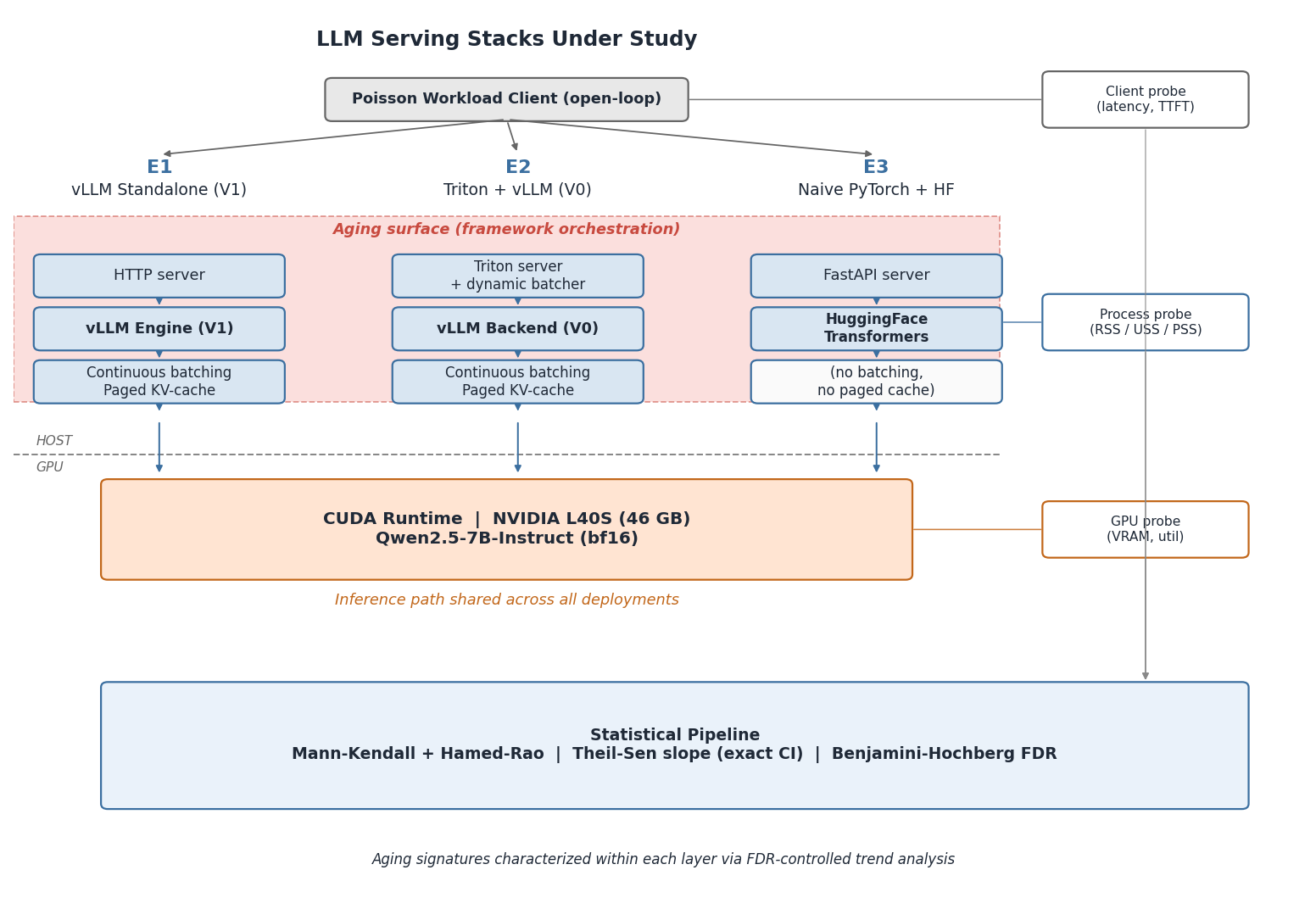}
\caption{The three LLM serving stacks studied. The model and GPU are shared across configurations; only the orchestration layers (the \emph{aging surface}) differ. Monitoring captures client-, process-, and GPU-side state in parallel, feeding a single statistical pipeline.}
\label{fig:arch}
\end{figure}

Our goal is to determine whether modern LLM serving engines on the GPU age over multi-hour timescales, and if so, where in the deployment stack the aging surface lies. This sets four design objectives: the campaign must compare practical deployment configurations under a controlled but realistic workload; monitor enough of the system to localize aging if present; use a statistical pipeline robust to the autocorrelation of monitoring traces; and remain reproducible end to end. Figure~\ref{fig:arch} summarizes the design: a shared inference path on the GPU, three host orchestration stacks, and monitoring probes that feed a common statistical pipeline. The rest of this section turns these objectives into concrete choices.

\textbf{Workload and stress regime.}
We drive each engine with a single open-loop Poisson client.
Prompts are drawn from a corpus of 3000 arXiv abstracts and
titles, concatenated to log-normal target lengths between 256 and
7500 tokens (median 1500); output length is log-normal with
median 200 and maximum 1500 tokens, and streaming is enabled with
probability 0.7. A short throughput calibration then fixes the
load: we raise the request rate until throughput stops growing,
which gives each engine its mechanical ceiling (i.e., its maximum
sustainable throughput), and we set the aging target rate at
85\% of it. We use the ceiling rather than the highest rate at
which the client does not yet drop, because the ceiling is a
property of the engine itself, while the no-drop rate is partly
an artefact of the client-side concurrency cap.

\textbf{Engines and ablations.}
The primary comparison spans three deployment configurations,
all serving the same model (Qwen2.5-7B-Instruct, bf16) on the
same hardware (NVIDIA L40S GPUs, 46~GB VRAM each): vLLM standalone
(E1), NVIDIA Triton wrapping vLLM (E2), and a naive PyTorch with
HuggingFace transformers (E3). Each runs for 36~hours, and the
three engines of a slot run concurrently, one per GPU, so they
share the host while each owns a separate device.
Table~\ref{tab:campaign} summarizes the full campaign. Two
targeted ablations complement the comparison. E3b re-runs the
naive baseline at low offered load to separate the load regime
from the framework's intrinsic behavior. A1 and A2 complete a
2$\times$2 factorial that crosses the two vLLM engine generations
(the legacy V0 and the redesigned V1) with the two hosting layers
(standalone server and Triton wrapper). The six runs together
amount to 216 hours of continuous engine operation.

\begin{table}[h]
\centering
\caption{Experimental campaign: six 36-hour aging runs.}
\label{tab:campaign}
\small
\begin{tabular}{@{}llllr@{}}
\toprule
ID  & Engine        & Hosting     & vLLM ver. & RPS target \\
\midrule
E1  & vLLM V1       & standalone  & 0.21.0    & 2.55       \\
E2  & vLLM V0       & Triton 25.09 & 0.10.1.1+nv & 2.17    \\
E3  & PyTorch+HF    & FastAPI     & n/a       & 0.17       \\
\midrule
E3b & PyTorch+HF    & FastAPI     & n/a       & 0.05       \\
A1  & vLLM V0       & standalone  & 0.7.3     & 0.80       \\
A2  & vLLM V1       & Triton 25.09 & 0.10.1.1+nv & 1.75    \\
\bottomrule
\end{tabular}
\end{table}

\textbf{Monitoring.}
Four streams of measurements run in parallel for the full
duration of each run, summarized in Table~\ref{tab:monitoring}. A
GPU monitor samples device-side state at 1\,Hz: VRAM occupancy,
utilization, temperature, power, clock frequencies, and ECC
counters. A process monitor samples the engine process every 5
seconds. The memory metrics we report in
Section~\ref{sec:results} are USS (the process-private resident
memory), RSS (the total physical memory used by the process), and
VMS (the virtual address space); we treat USS as the primary leak
indicator, since it counts only pages private to the process. The
same monitor also tracks thread and file-descriptor counts, CPU
usage, context-switch rates, and I/O rates. A system monitor
samples host-level state every 5 seconds: memory and swap usage,
load average, CPU usage, and file-descriptor counts. A client
logger records one row per request, with timestamps, status,
token counts, and end-to-end latency. The catalog spans about 34
indicators per run.

\begin{table}[ht]
\centering
\caption{Monitoring streams. USS, RSS, VMS, and PSS are different
accountings of process memory (see text).}
\label{tab:monitoring}
\small
\begin{tabular}{@{}lllp{0.48\linewidth}@{}}
\toprule
Stream & Source & Period & Indicators \\
\midrule
GPU     & NVML   & 1~s     & VRAM, util, temp, power, clocks, ECC \\
Process & psutil & 5~s     & USS, RSS, VMS, PSS, threads, FDs, CPU, ctx-switches, I/O rates \\
System  & psutil & 5~s     & mem, swap, load, CPU, FDs \\
Client  & log    & per req & lat, TTFT, tokens, status \\
\bottomrule
\end{tabular}
\end{table}

\textbf{Statistical Analysis.}
Monitoring traces are strongly autocorrelated: the resident
memory of a process at time $t$ is almost identical to its value
at $t-1$, and naive tests that treat samples as independent
produce drastically deflated p-values. We therefore run two
complementary analyses on each indicator. A non-parametric trend
test, Mann-Kendall with the Hamed-Rao
correction~\cite{mann1945,hamed1998modified}, inflates the test
variance by an effective-sample-size factor from the empirical
autocorrelation function and answers whether a monotonic trend
exists. A non-parametric slope estimator, Theil-Sen with an exact
confidence interval from the order statistics of pairwise
slopes~\cite{sen1968,theil1950,hollander1999nonparametric},
inflates the variance by the lag-1 AR(1) factor
$(1+\rho)/(1-\rho)$ and gives the magnitude and its uncertainty.
We declare a trend significant only when the two agree:
Mann-Kendall rejects the null and the 95\% Theil-Sen interval
excludes zero. Across about 34 indicators and six runs, roughly
200 trend tests, we control the false discovery rate at $q=0.10$
with Benjamini-Hochberg~\cite{benjamini1995fdr}. The indicator
list is fixed by the monitor schemas before any long-run data is
examined, a de facto pre-registration of the analysis, and
cumulative psutil counters (context switches, I/O bytes) are
replaced by their per-second rates, since they trend
monotonically by construction and would yield trivially
significant ``aging''.

\section{Results}
\label{sec:results}
We report the results in four steps. First, as a production
operator would see it, through latency and throughput, the
engines show almost no aging over 36 hours. Second, inside the
engine process we find a small but statistically significant
memory leak in every deployment, and the order is surprising: the
naive baseline is not the cleanest, the optimized standalone
engine is. Third, a low-load run shows the leak is not caused by
heavy load, because it is still there when the same engine runs
far below its limit. Fourth, the 2$\times$2 factorial, which
crosses engine generation with hosting layer, shows that the leak
rate depends on the whole deployment, not on any single part: it
spans about two orders of magnitude across the four cells, and is
largest when Triton wraps the legacy V0 engine.

\subsection{Campaign overview}
\label{sec:campaign-stability}
Since an aging signal can only build up under load, we first
report the aggregate operational metrics of the six runs, to
confirm that the intended stress regime was reached and held for
the full 36~hours. Table~\ref{tab:aggregates} summarizes them.
The three primary engines (E1, E2, E3) tracked their target rates
within 0.4\% and processed between 22{,}532 and 329{,}327
requests each. Their drop rates match the regime they were set
for: vLLM standalone and Triton ran near their ceiling with
negligible drop ($\leq$0.01\%), while the naive PyTorch+HF
baseline ran saturated, dropping 15.6\% of requests. The low-load
ablation (E3b) confirms the sub-saturated regime, with no drops
and a p50 latency 60$\times$ lower than E3. The V0/V1 ablations
(A1, A2) ran without drops at their target rates. No run
collapsed, restarted, or showed any discontinuity. The six runs
together deliver 216 hours of continuous operation under load, so
the trends that follow can be read as aging under genuine stress,
not drift in an idle system.
\begin{table}[h]
\centering
\caption{Aggregate metrics of the six 36-hour aging runs.}
\label{tab:aggregates}
\small
\begin{tabular}{@{}llrrrrr@{}}
\toprule
ID  & Config       & Req   & Drop\% & p50   & p99   & tok/s\footnotemark \\
\midrule
E1  & vLLM V1 std  & 329k  & 0.01   & 7.93  & 51.92 & 686   \\
E2  & Triton+V0    & 280k  & 0.01   & 8.70  & 59.18 & n/a   \\
E3  & PyT+HF sat   & 23k   & 15.60  & 405   & 536   & 40    \\
E3b & PyT+HF low   & 6k    & 0.00   & 6.59  & 43.05 & 13    \\
A1  & vLLM V0 std  & 103k  & 0.00   & 5.17  & 34.27 & 211   \\
A2  & Triton+V1    & 227k  & 0.00   & 6.32  & 42.82 & n/a   \\
\bottomrule
\end{tabular}
\end{table}
\footnotetext{Output throughput is computed from the per-request
output token counts in the client log. The Triton
OpenAI-compatible client path does not report these counts, so
output throughput is \emph{n/a} for the Triton-hosted cells
(E2 and A2).}

\subsection{Client-side analysis}
\label{sec:client-stationarity}
If aging is occurring, users are likely to notice it through slower responses, reduced throughput, or more errors. We therefore look for trends in these client-side metrics across the six 36-hour runs.

The answer is almost entirely negative. Across the three primary
engines (E1, E2, E3) and the two ablation cells (A1, A2),
end-to-end latency (p50, p95, p99), time-to-first-token (p50,
p99), output throughput, and request drop rate are stationary,
with Theil-Sen confidence intervals that include zero. The only
exceptions are a few small trends on E1: its p50 latency rises by
about 5~ms/h and its p50 time-to-first-token by a fraction of a
millisecond per hour, while its output throughput rises slightly
over the same window. These are statistically significant but
tiny, on the order of two percent of the baseline over 36 hours,
and the throughput moves the wrong way for aging, so none of them
is degradation an operator would notice.

Drop rate is also stationary within each run. The naive baseline
E3 drops about 15.6\% of requests throughout, but this is a steady
capacity ceiling, not a growing loss: the rate does not trend over
the 36 hours (Section~\ref{sec:regime-ablation}).

The takeaway from a black-box perspective is that, for all
practical purposes, none of the engines degrades visibly to a
production operator over 36 hours of sustained high load. As the
next section shows, this stability is only skin-deep: once we look
inside the engine process, a different picture emerges.

\subsection{Process-side analysis}
\label{sec:process-leaks}
Client-side metrics may remain stable even when memory accumulates inside the engine. We therefore analyze the engine's memory footprint over time using USS, RSS, and PSS (Section~\ref{sec:methodology}), with particular attention to USS, the memory private to the process.
The result is positive for all three deployments. vLLM standalone
(E1), Triton-wrapped vLLM (E2), and the naive PyTorch+HuggingFace
server (E3) each show a small but statistically significant,
steadily rising memory trend after FDR correction. The
autocorrelation is near one ($\rho \approx 0.99$), consistent with
slow, roughly linear growth over the full 36-hour window, and the
USS and RSS slopes agree to within rounding, so the growth is in
pages private to the process, not in mappings shared between
processes.

Table~\ref{tab:primary-slopes} gives the slopes and their 95\%
confidence intervals, and the order is the surprise of the
campaign. The leak is small everywhere, tens of kilobytes per
hour. The naive PyTorch baseline, the textbook example of what
production teams avoid, is \emph{not} the cleanest: it sits in the
middle at $+31$~KB/h. The cleanest is vLLM standalone (E1) at
$+1.8$~KB/h, an engine built specifically to optimize serving.
The heaviest is Triton-wrapped vLLM (E2) at $+157$~KB/h. Over a
30-day deployment these rates add up to roughly 1, 22, and
110~MB of private memory per replica, well short of an
operational concern on monthly timescales.

\begin{table}[h]
\centering
\caption{Process-private memory leak rate in the three primary
deployments. Slopes are Theil-Sen USS estimates per hour with
95\% confidence intervals.}
\label{tab:primary-slopes}
\small
\begin{tabular}{@{}llrr@{}}
\toprule
ID & Deployment             & USS slope/h     & 95\% CI         \\
\midrule
E1 & vLLM standalone (V1)   & +1.8 KB         & [0.7, 23.8] KB  \\
E2 & Triton + vLLM (V0)     & +157 KB         & [33, 228] KB    \\
E3 & PyTorch + HF naive     & +31 KB          & [21, 88] KB     \\
\bottomrule
\end{tabular}
\end{table}

The relative ordering of the leak rates is as informative as the leaks themselves. All three deployments use the same model weights and execute the same inference computations. What differs is the runtime environment around the model. If the leak originated from the inference path, we would expect similar leak rates across all deployments. Instead, the observed rates differ by nearly two orders of magnitude. This strongly suggests that the source of the leak lies in the serving infrastructure rather than in the model execution itself.

The results also indicate that leak severity is not simply related to runtime complexity. The standalone engine with the largest software stack exhibits the lowest leak rate, while the Triton deployment built on the legacy engine exhibits the highest. Memory aging therefore appears to depend on the specific runtime implementation and deployment configuration. The factorial analysis in Section~\ref{sec:v0v1-ablation} explores this relationship in more detail.

The remaining process metrics tell a consistent story. CPU utilization, thread count, file-descriptor count, context-switch rate, and I/O activity remain largely stable over time. Although the trend test identifies a few statistically significant changes, their magnitude is negligible. At this timescale, aging manifests as a memory issue rather than as scheduler drift, descriptor leakage, or I/O degradation.

The GPU metrics show a similar pattern. VRAM usage exhibits no significant trend in any deployment according to our decision rule. This result is somewhat surprising because we expected heterogeneous request sizes to gradually increase fragmentation in the PyTorch caching allocator. However, no such effect is visible during the 36-hour observation period.

\subsection{Effect of workload regime}
\label{sec:regime-ablation}
One possible explanation for the leak observed in Section~\ref{sec:process-leaks} is that it is specific to the saturated operating regime of the PyTorch+HF server, where only a limited set of execution paths may be exercised. To test this hypothesis, we repeated the experiment at $0.05$~RPS (E3b), well below saturation.
The leak persists under this lighter workload and is in fact larger, increasing from approximately $+31$~KB/h in the saturated configuration to $+103$KB/h in the sub-saturated one. This result indicates that the leak is not caused by queueing effects or saturation-related behavior. Instead, it appears to be intrinsic to the runtime environment. The higher leak rate at lower load also suggests that memory growth is driven primarily by elapsed time rather than by the number of processed requests.
Both PyTorch deployments exhibit another interesting behavior. Virtual memory grows much faster than resident memory. In E3b, the virtual address space increases by about $+127$MB/h, corresponding to roughly 7.5GB over the 36-hour run, while resident memory increases by only about 20MB. At the same time, thread counts and file-descriptor counts remain stable. These observations rule out thread or descriptor leakage and instead point to ongoing reservation of anonymous address space, which is consistent with the host-side management performed by the CUDA caching allocator.

Whether this virtual-memory growth eventually translates into a measurable resident-memory cost remains unclear. The duration of our experiments is sufficient to expose the trend, but not to determine its long-term consequences.

\subsection{Engine generation and hosting layer}
\label{sec:v0v1-ablation}

\begin{figure}[h]
\centering
\includegraphics[width=\linewidth]{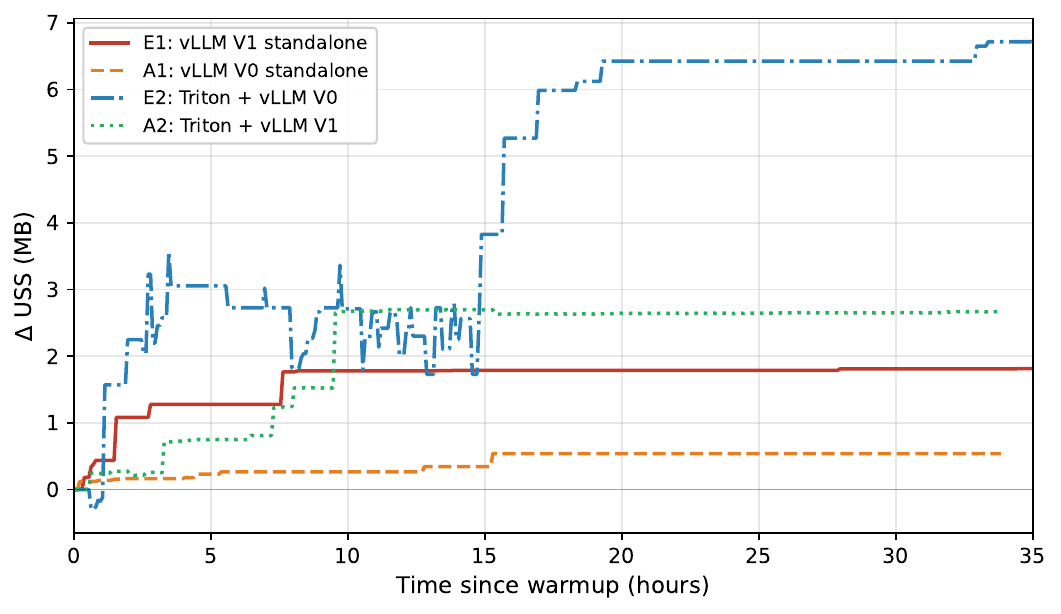}
\caption{Process-private memory growth (USS) over the 36-hour
window for the four factorial configurations, normalized to the
post-warmup baseline. Only E2 (Triton + V0) exhibits step-wise
growth; the other three configurations grow smoothly.}
\label{fig:rss-combined}
\end{figure}

Sections~\ref{sec:process-leaks} and~\ref{sec:regime-ablation}
showed that memory aging originates in the serving runtime rather
than in the inference path itself. However, E1 and E2 differ
along several dimensions, including vLLM version, engine
generation (V1 in E1 and the legacy V0 in E2), and hosting layer
(standalone versus Triton). To disentangle these effects, we add
two configurations that complete a $2\times2$ factorial design
across engine generation and hosting layer: vLLM standalone on V0
(A1) and Triton on V1 (A2). Table~\ref{tab:v0v1-factorial}
summarizes the four configurations, while Figure~\ref{fig:rss-combined} shows their memory trajectories.

\begin{table}[h]
\centering
\caption{$2\times2$ factorial on USS leak rate.}
\label{tab:v0v1-factorial}
\small
\begin{tabular}{@{}lrr@{}}
\toprule
                & V0 engine          & V1 engine          \\
\midrule
Standalone      & A1: $+13$~KB/h     & E1: $+1.8$~KB/h    \\
Triton-wrapped  & E2: $+157$~KB/h    & A2: $+5.9$~KB/h    \\
\bottomrule
\end{tabular}
\end{table}

The pattern is clear across all four configurations. The legacy
V0 engine leaks more than V1 under both hosting layers. The gap
is approximately $7\times$ in standalone mode (A1 versus E1) and
about $27\times$ under Triton (E2 versus A2). Triton also
amplifies the leak for both engines, with a much stronger effect
for V0. The cleanest configuration is standalone V1 (E1), with a
leak rate of only $+1.8$KB/h, corresponding to roughly 1MB
over 30 days. The worst case is Triton running V0 (E2), whose
leak rate exceeds all others by a wide margin. Overall, the
results indicate that memory aging depends on the combination of
engine generation and hosting layer rather than on either factor
alone. Across the four configurations, leak rates span nearly two
orders of magnitude.

One detail deserves attention. In Figure\ref{fig:rss-combined}, E1 rises during the first few hours and then stabilizes, whereas A1 continues to grow slowly
throughout the experiment. As a result, E1 does not exhibit the
smallest cumulative growth despite having the lowest sustained
leak rate. The slopes reported in
Table\ref{tab:v0v1-factorial} capture long-term growth, whereas
the initial settling phase of E1 does not.

The trajectories also reveal a qualitative difference that is not
captured by the leak-rate estimates. Three configurations (E1,
A1, and A2) exhibit smooth growth, consistent with a gradual and
fine-grained leak. E2 behaves differently. Its memory footprint
remains stable for extended periods and then increases through
discrete jumps of several megabytes. To characterize this behavior, we examine the distribution of
memory increments over consecutive observation windows. For E2,
the increments are concentrated and heavy-tailed. The largest one
percent of increments approach 1.8MB, and resident and virtual
memory increase together, with a lag-zero correlation of
$\rho \approx 0.83$. In the other three configurations, memory
increments remain near zero and below one megabyte
($\rho < 0.35$), which is consistent with gradual drift rather
than step-wise growth.

This E2 signature is particularly distinctive. RSS and VMS grow
together through discrete allocation events and are never fully
released. One plausible explanation is that the engine
periodically maps new memory regions from the operating system
and retains them over time. Confirming the underlying mechanism
would require heap-level profiling, which is outside the scope of
this study. Nevertheless, this behavior is the strongest clue
emerging from the experimental campaign and the most plausible
candidate for unbounded growth over multi-day timescales.

The GPU results tell a different story. None of the four
configurations shows a significant VRAM trend over the 36-hour
observation window, including standalone V0 (A1). Once the model
is loaded, VRAM occupancy remains essentially stable.

One caveat should be noted. The four configurations span three
different vLLM versions (V0 at 0.7.3, the Triton configurations
at 0.10.1.1, and standalone V1 at 0.21.0), because recent vLLM
releases no longer allow the legacy engine to be selected
explicitly. As a result, the effect of engine generation cannot
be fully separated from version differences. Removing this
confounding factor would require maintaining a custom fork. The
observed trends are unlikely to be explained solely by version
drift given the magnitude of the differences, but a precise
attribution remains an open question.

The main conclusion is that memory aging in modern LLM serving is
a property of the complete deployment stack. The legacy V0 engine
leaks more than V1, Triton amplifies the leak, and the
combination of Triton and V0 produces both the highest leak rate
and the only step-wise growth pattern. This distinctive signature
appears to originate within the V0 execution path and represents
a natural target for future reliability investigations.
\section{Threats to Validity}
\label{sec:threats}

We discuss the threats with the greatest impact on our findings,
roughly in order of severity. One of them, version drift in the
factorial experiment, was already noted in
Section~\ref{sec:v0v1-ablation}.

\textbf{Single-run design.}
Each configuration is observed in a single 36-hour run, with no
replicates. The reported slopes and confidence intervals capture
within-run variation, but not run-to-run variability, which would
require independent repetitions. The strongest part of our
findings is the existence of aging and the relative ordering of
the deployments, since the leak rates differ by nearly two orders
of magnitude. The exact slopes should instead be interpreted as
point estimates from a single run. Replication with
$n \geq 3$ runs per configuration is the most immediate next
step.

\textbf{Version drift across the factorial.}
The $2\times2$ factorial spans three different vLLM versions
across its four cells (V0 at 0.7.3, the Triton cells at 0.10.1.1,
and standalone V1 at 0.21.0), because recent releases no longer
allow the legacy engine to be selected explicitly. The
engine-generation effect is therefore partially confounded with
version drift. Removing this confounding factor would require
maintaining a custom fork. Given the size of the observed gaps, it
is unlikely that the qualitative findings are solely due to
version differences, but a precise attribution remains open.

\textbf{Parallel multi-tenant topology.}
The three engines in each slot run concurrently on separate GPUs
of the same host and therefore share CPU, memory, and I/O
resources. This reflects a realistic multi-tenant deployment, but
it also means that the measured leak rates include host-level
contention. The same engines running on dedicated hosts might
exhibit somewhat different rates. We therefore treat contention
as part of the operating environment rather than as experimental
noise.

\textbf{Workload assumption.}
Each engine is driven by a stationary open-loop Poisson workload
using synthetic prompts. This provides a controlled and
reproducible setup, but it does not capture the burstiness,
temporal locality, or prompt repetition often seen in production
traffic. Some aging mechanisms may emerge under such workloads,
while others may become less pronounced. Evaluating the same
systems under production traces or workloads with varying
burstiness and repetition would help assess the generality of the
observed signatures.

\textbf{Single hardware and single model.}
The campaign uses one GPU class (NVIDIA L40S) and one model
(Qwen2.5-7B-Instruct). Other GPUs (A100, H100, B200) and larger
or architecturally different models may exhibit different memory
management behavior. We expect the qualitative findings to carry
over, namely that every deployment exhibits a small
process-private leak, that aging originates in the runtime rather
than the inference path, that deployment choices influence leak
severity, and that the step-wise pattern is unique to Triton+V0.
The quantitative results, however, will need to be re-measured.

\section{Final Remarks}
\label{sec:discussion}

We studied software aging in GPU-based LLM serving through a
216-hour experimental campaign covering six deployment
configurations and analyzed the results using a statistical
pipeline designed for autocorrelated monitoring data. Three main findings emerge. First, software aging is present in
this domain. Every deployment exhibits a small but statistically
significant process-private memory leak over a 36-hour window.
Second, aging is not a property of the model or inference path
alone. Deployments running the same model and serving the same
workload can differ by nearly two orders of magnitude in leak
rate, showing that the runtime and hosting environment play a
central role. Third, not all leaks look the same. Most
deployments exhibit smooth memory growth, whereas Triton+V0 shows
a distinctive step-wise pattern in RSS and VMS, suggesting a
different underlying mechanism. Notably, none of these effects is
visible through client-side metrics over the timescale of our
experiments. For practitioners, the results highlight that monitoring LLM
serving systems cannot rely solely on latency, throughput, or
error rates. Memory-related aging may remain invisible to users
while steadily accumulating inside the serving stack. More
generally, deployment choices have reliability implications that
are not captured by short-term performance benchmarks.

Several directions remain open. Replicated experiments would
quantify run-to-run variability, while extending the study to
other GPU platforms and larger models would test the generality
of the observed patterns. Longer observation windows could reveal
late-onset aging effects, and targeted heap profiling of the V0
engine could identify the source of the step-wise accumulation.
Finally, understanding whether specific request patterns can
trigger or accelerate these memory-growth mechanisms remains an
interesting question for future work.

\bibliographystyle{IEEEtran}
\bibliography{refs}

\end{document}